\documentclass[aps,prl,twocolumn,showpacs,groupedaddress,amsmath,amssymb]{revtex4}
\usepackage{graphicx}
\usepackage{color}
\usepackage{bm}
\usepackage{latexsym}

\begin{document}
\title{Perfect Impedance-Matched Isolators and Unidirectional Absorbers}
\author{J. M. Lee$^{1}$, Z. Lin$^{1}$, H. Ramezani$^{1}$, F. M. Ellis$^{1}$, V. Kovanis$^{2}$, I. Vitebskiy$^{2}$, T. Kottos$^{1}$}
\affiliation{$^1$Department of Physics, Wesleyan University, Middletown, CT-06459, USA}
\affiliation{$^2$Air Force Research Laboratory, Sensors Directorate, Wright Patterson AFB, OH 45433 USA}
\date{\today}
\begin{abstract}
A broad-band reflectionless channel which supports unidirectional wave propagation originating from the interplay between gyrotropic 
elements and symmetrically placed gain and loss constituents is proposed. Interchange of the active elements together with a gyrotropic 
inversion turns the same structure to a unidirectional absorber where incoming waves from a specific direction are annihilated. When 
disorder is introduced asymmetric Anderson localization is found. Realizations of such multi-functional architectures in the frame of electronic 
and photonic circuitry are discussed. 
\end{abstract}

\pacs{42.25.Bs, 42.25.Hz, 03.65.Nk }

\maketitle
 
Understanding and controlling the direction of wave propagation is at the heart of many fundamental problems of physics while it is 
of great relevance to engineering. In the latter case the challenge is to design on-chip integrated devices that control energy flow in 
different spatial directions. Along these lines, the creation of novel classes of integrated photonic, electronic, acoustic or thermal 
diodes is of great interest.  Such unidirectional elements constitute the basic building blocks for a variety of transport-based devices 
like rectifiers, circulators, pumps, coherent perfect absorbers, molecular switches and transistors \cite{K05,ST91,WCGSC11,LC11}. 

The idea was originally implemented in the electronics framework, with the construction of electrical diodes that were able to rectify 
the current flux \cite{K05}. This significant revolution motivated researchers to investigate the possibility of implementing the notion 
of ``diode action" to other areas of physics ranging from  thermal \cite{TPC02} and acoustic transport \cite{NDHJ05} to optics \cite{ST91}. 
Specifically in optics, unidirectional elements rely almost exclusively on magneto-optical (Faraday) effects. However, at optical 
frequencies all nonreciprocal effects are very weak resulting in prohibitively large size of most nonreciprocal devices.  Alternative 
proposals include the creation of optical diodes based on nonlinearities \cite{GAPF01,SDBB94,B08,LC11} . These schemes however suffer 
from serious drawbacks and limitations since the rectification depends on the level of incident power or/and whether the second harmonic 
of the fundamental wave is transmitted or not. Obviously, in this later case the outgoing signal does not have the same characteristics 
as the incident one. Moreover non-linearities can result in enhanced reflection, which significantly compromises the diode
performance.

\begin{figure}
   \includegraphics[width=.85\linewidth, angle=0]{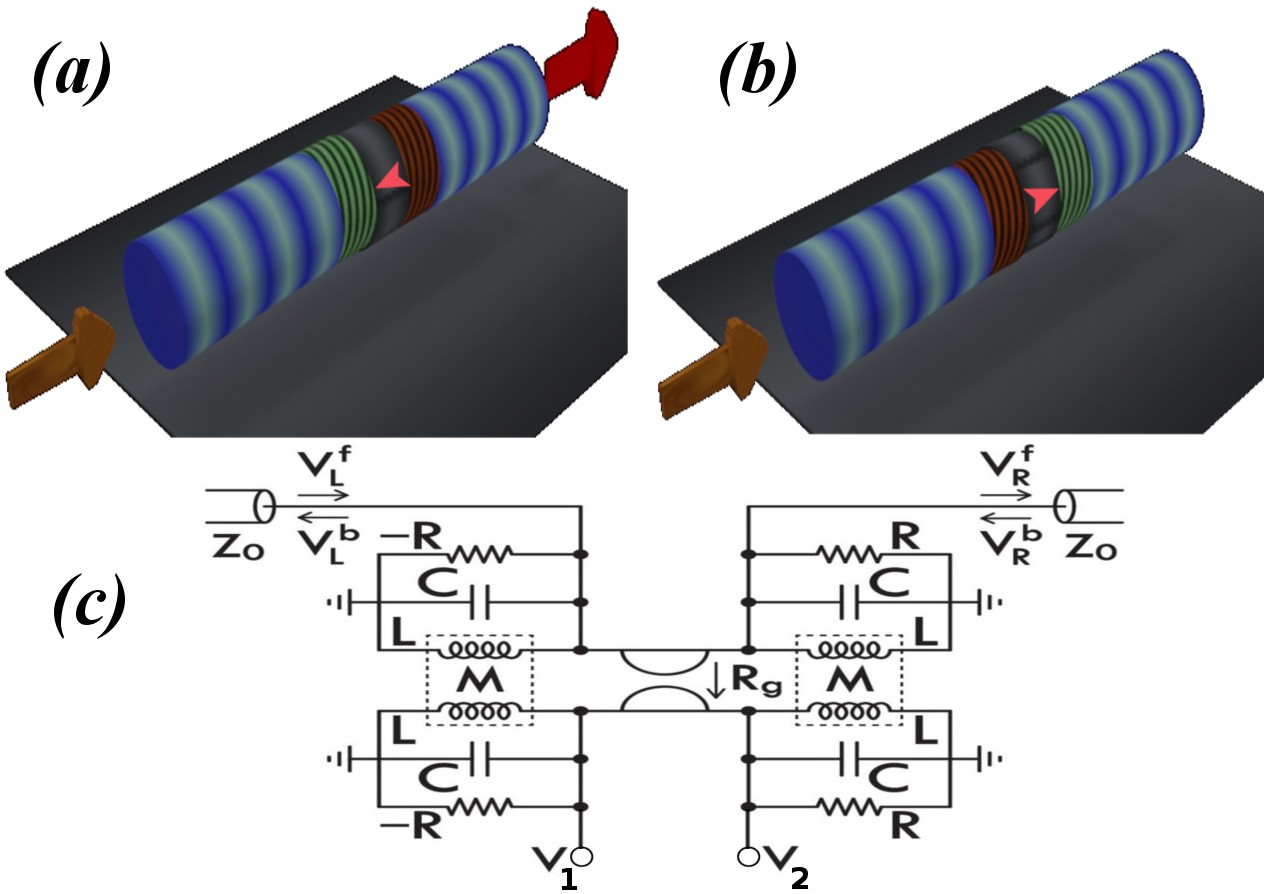}
    \caption{(Color online) A ${\cal P{\tilde T}}$-symmetric photonic structure. Red and green layers indicate gain and loss. The grey layer indicates 
a magneto-optical material. The orange arrow indicates the direction of the magnetic field (gyrotropy). Light/dark blue layers indicate a Bragg grating 
which is missaligned with the active layers. (a) A perfect impedance-matched isolator where reflection is vanished. (b) The same structure
can also act as a unidirectional absorber if we exchange the gain/loss layers and invert the gyrotropy. (c) The equivalent electronic circuit. $R_g$
indicate the gyrator.
\label{fig1}}
\end{figure}

In this Letter, we propose a physical setting (see Fig. \ref{fig1}) which acts as a perfect one-way valve i.e. a channel along which waves 
propagate in only a single direction, with zero reflection, a {\it perfect impedance matching isolator}. This is achieved by employing an 
interface between a gyrotropic element and two active constituents, one with gain and another one with a balanced amount of loss. 
In contrast to standard parity-time (${\cal PT}$)-symmetric systems, first proposed in the framework of optics by Christodoulides and 
colleagues \cite{Makris}, our structure shows a new type of {\it generalized} ${\cal P{\tilde T}}$-symmetry which allows for non-reciprocal 
transport. Furthermore, an exchange of the active elements together with an inversion of the gyrotropy leads to {\it unidirectional absorption} 
where an incoming wave entering the structure from one side is completely annihilated.  
These features can be observed over a broad range of frequencies when 1D chains of such elements are considered. Finally, the presence 
of imperfections results in destructive interferences that are sensitive to the direction of propagation, a phenomenon refer to as 
{\it asymmetric Anderson localization}. Below we present realizations of such multi-functional architectures, both in the frame of 
photonics and electronics circuitry (EC). The later framework has been proven recently \cite{SLZEK11} extremely useful for the investigation 
of the transport properties of ${\cal PT}$-symmetric systems.


The photonic structures  that we consider consist of a central magneto-optics layer sandwiched  between two equally balanced gain 
and loss birefringent layers which allow for coupling between the $x-y$ polarizations (see Fig. \ref{fig1}a). The role of the magnetic 
element is to induce magnetic non-reciprocity which is associated with the breaking of time reversal symmetry. Breaking time-reversal 
symmetry is not sufficient to obtain non-reciprocal transport - the theory of magnetic groups shows that the absence of space inversion 
symmetries is also required. This is achieved with the use of the two active birefringent layers \cite{RLKKKV12} and an anisotropic Bragg 
mirror which has missalignment with the active layers; therefore it does not allow for coupling between the two polarization channels.

The equivalent EC (Fig. \ref{fig1}b) comprised of two pairs of mutually inductive, $M = \mu L$, coupled $LC$ oscillators, one with 
amplification (left column of Fig.~\ref{fig1}b) and the other with equivalent attenuation (right column of Fig.~\ref{fig1}b). The loss imposed 
on the right half of the structure is a standard resistor $R$. Gain imposed on the left half of the EC, symbolized by -$R$, is implemented 
with a negative impedance converter (NIC). The NIC gain is trimmed to oppositely match the value of R used on the loss side, setting 
the gain and loss parameter $\gamma = R^{-1}\sqrt{{L\over C}}$. The uncoupled frequency of each resonator is $\omega_0=1/\sqrt{LC}$. 
The pairs are coupled with each other via a gyrator \cite{T48}. This element immitates the role of the magnetic layer used in the photonic 
set-up. The gyrator is a lossless two-port network component with an antisymmetric impedance matrix $Z_g=-Z_g^T$ connecting the 
input and output voltages ${\bf  V}\equiv (V_{n},V_{m})^T$ and currents ${\bf I}\equiv (I_{n},I_{m})^T$ associated with ports $n$ and $m$ as
\begin{equation}
\label{gyrator}
{\bf V}=Z_g {\bf I};  \quad  (Z_g)_{n,m}=(-1)^{n}(\delta_{n,m}-1) R_g
\end{equation}
where $R_g =\beta^{-1}\sqrt{{L\over C}}$ and $\beta$ is a dimensionless conductance. Eq. (\ref{gyrator}) is invariant under a {\it generalized} 
time reversal operator $\tilde{\cal T}$ which performs a combined time-inversion ($t\rightarrow -t$) together with a transposition of $Z_g$. 
The anti-linear operator $\tilde {\cal T}$ is equivalent to the transformation $t\rightarrow-t; \beta\rightarrow -\beta$. Despite the fact that
a gyrator breaks the time-reversal symmetry, it does not alone leads to non-reciprocal transport  (in a similar manner that a magnetic
layer is not sufficient to create asymmetric transport in the optics framework).

Finally, the two (upper/lower) propagation channels that are supported by the EC structure of Fig. \ref{fig1}b imitate the $x/y$ polarization 
channels of the photonic set-up.  In the latter case, the BG act as a polarization filter for incident waves with frequencies chosen from the
pseudo-gaps. These pseudo-gaps are a consequence of the anisotropic refraction index along the $x$ and $y$ polarization channels. The 
analogous effect can be achieved in the EC case by coupling only the upper (or lower) channel to a transmission line (TL) with impedance $Z_0$. 

Although below we exploit the simplicity of the electronic (lump) circuitry framework in order to guide the physical intuition and derive 
theoretical expressions for the transport characteristics of these structures, we will always validate our results in the realm of photonic 
circuitry via detail simulations.


At any point along a TL, the current and voltage determine the amplitudes of the right and left traveling wave components.  The forward 
$V_{L/R}^{+}$ and backward $V_{L/R}^{-}$ wave amplitudes, and $V_{L/R}$ and $I_{L/R}$ the voltage and current at the left (L) or right 
(R) TL-EC contacts satisfy the continuity relations
\begin{equation}
\label{scatstates}
V_{L/R}=V_{L/R}^{+}+V_{L/R}^{-};\quad I_{L/R}=\left[V_{L/R}^{+}-V_{L/R}^{-}\right]/Z_0. 
\end{equation}
Note that with this convention, a positive lead current flows out of the right
circuit, but into the left circuit, and that the reflection amplitudes 
for left or right incident waves are defined as $r_{L,R}\equiv {V_{L,R}^{\mp}/V_{L,R}^{\pm}}$ and $t_{L,R}\equiv {V_{R,L}^{\pm}/V_{L,R}^{\pm}}$ 
respectively.

Application of the first and second Kirchoff's laws at the TL-EC contacts allow us to find the current/voltage wave amplitudes at the left and right
contact. We get
\begin{align}
i [\gamma V_L-\beta(V_1 - V_2)] -  \omega V_L + { V_L - \mu V_1 \over \omega {\bar m}} &=-i\eta Z_0 I_L  \notag \\
i [\gamma V_R-\beta(V_1 - V_2)] + \omega V_R - { V_R - \mu V_2 \over \omega {\bar m}} &=-i \eta Z_0 I_R \notag \\
i [\gamma V_1+\beta(V_L - V_R)] - \omega  V_1+ {V_1 - \mu V_L  \over \omega {\bar m}} &=0   \notag \\
i [\gamma V_2+\beta(V_L - V_R)] + \omega  V_2- {V_2 - \mu V_R \over \omega  {\bar m}} &=0  
\label{Kirchoff}
\end{align}
where ${\bar m}=1-\mu^2$, $V_{1,2}$ is the voltage at the lower channels (see Fig.~\ref{fig1}a), $\gamma$ is the gain/loss parameter, and $\eta=\sqrt{L/C}/Z_0$ 
is the dimensionless TL impedance. Here, the dimensionless wave frequency $\omega$ is in units of $\omega_0$. Equations (\ref{Kirchoff})
are invariant under a combined parity ${\cal P}$ (i.e. $L\leftrightarrow R, 1\leftrightarrow 2$) and $\tilde{\cal T}$ (i.e. $i\leftrightarrow -i, 
\beta\leftrightarrow -\beta$) transformation (we recall that the time-reversal $t\rightarrow -t$ is equivalent to conjugation). 

For the EC structure of Fig. \ref{fig1}b, the transmission and reflection amplitudes in terms of the gyrotropic and dissipative conductances 
$\beta$ and $\gamma$ are derived using Eqs. (\ref{Kirchoff})
\begin{eqnarray}
t_R &=  {f(\omega;\beta,\gamma)\over h(\omega;\beta,\gamma)};\quad
t_L &=  {f(\omega;\beta,-\gamma)\over h(\omega;\beta,\gamma)}\nonumber\\
r_R &=  {g(\omega;\beta,-\gamma)\over ih(\omega;\beta,\gamma)};\quad
r_L &=  {g(\omega;\beta,\gamma)\over ih(\omega;\beta,\gamma)}
\label{exact}
\end{eqnarray}
where the functions $f(...), g(...)$ and $h(...)$ are given at the supplement and satisfy the symmetry relations $f(\omega;\beta,\gamma)
=f(\omega;-\beta,-\gamma)$, $g(\omega;\beta,\gamma) = g(\omega;-\beta,\gamma)$ and $h(\omega; \beta,\gamma)=h(\omega;
|\beta|,|\gamma|)$. Consequently, these symmetry relations can be translated to the following relations for the transmission and reflection
amplitudes for any frequency $\omega$
\begin{equation}
\begin{array}{cccc}
t_R(\beta,\gamma)=&t_R(-\beta,-\gamma)&=t_L(-\beta,\gamma)&=t_L(\beta,-\gamma) \,\\
r_R(\beta,\gamma)=&r_R(-\beta,\gamma)&=r_L(\beta,-\gamma) &= r_L(-\beta,-\gamma).
\end{array}
\label{symmetries}
\end{equation}
The corresponding left (L) and right (R) transmitance and reflectance are then defined as $T_{L/R}\equiv 
|t_{L/R}|^2$ and $R_{L/R}\equiv |r_{L/R}|^2$ respectively.

The ratio between the left and right transmission takes the simple form
\begin{equation}
\label{ratio}
{t_L \over t_R} =  1 + \frac{2 \gamma \mu}{\beta - \gamma \mu - \beta  (1- \mu^2)\omega^2}.
\end{equation}
which establishes the nonreciprocal nature of ${\cal P{\tilde T}}$-symmetric transport. Note that $\gamma=0$ results in reciprocal transport 
irrespective of the value of $\beta$. The same conclusion is reached for $\beta=0$, corresponding to standard ${\cal PT}$-symmetric structures.  
Therefore we conclude that non-reciprocal transport is achieved due to the the interplay between gain/loss elements (i.e. $\gamma$) together
with gyrotropic constituent  $\beta$. 

Furthermore one can show that the left (right) transmission $t_{L(R)}$ and reflection $r_{L(R)}$ amplitudes satisfy the relations
\begin{equation}
 \operatorname{Re}[~t_L r_R^*~] = 0, \quad
 \operatorname{Re}[~t_R r_L^*~] = 0, \quad
 t_L t_R^* + r_R r_L^* = 1 
\label{conserv}
\end{equation}
which again implies that in general transport is non-reciprocal. In the photonic case the transmission and reflection amplitutes becomes 
$2\times 2$ matrices. In fact, Eqs. (\ref{conserv}) are consequences of generalized unitarity relations satisfied by the scattering matrix 
$S$ \cite{RLKKKV12,Sc12}.

\begin{figure}
   \includegraphics[width=1.0\linewidth, angle=0]{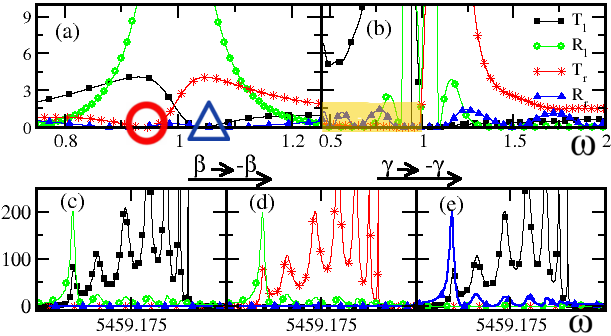}
    \caption{(Color online) (a)  Transmissions and reflections $T_{L/R};R_{L/R}$ of one EC unit for a left/right incident wave. The red circle 
indicates a UA while the blue triangle indicates an IMI frequency for a right incident wave; (b) A broad-band UA for a right incident wave
for a periodic 1D array of ${\cal N}=10$ EC units. We have checked (not shown) that changing $\beta\rightarrow -\beta$ we get an IMI 
for the same frequency window (see text); (c) A broad-band UA for an $x$-polarized right incident wave in the case of a periodic photonic 
structure; (d) Changing $\beta\rightarrow -\beta$ leads to an IMI behavior (see text) for the set-up of subfigure (c); (e) Additional change 
of $\gamma\rightarrow -\gamma$ leads to UA for a left incident wave. 
\label{fig2}}
\end{figure}

In Fig. \ref{fig2}a we report representative cases of the reflected and transmitted signals for a left and right incident waves. A striking feature of 
these plots is the fact that at specific frequencies $\omega_p$ the transmittance from one side can become zero, while from the other side 
is high.  At these specific values of ($\omega_p, \gamma_p, \beta_p$), the EC acts as a {\it perfect} isolator. For example, the requirement 
for zero transmittance for a left incoming wave $t_L=0$ leads to
\begin{equation}
 \omega_p(\gamma_p, \beta_p) = \sqrt{{\beta_p + \gamma_p \mu \over \beta_p  (1 - \mu^2)}} 
\label{ws}
\end{equation}
while the frequency where $t_R = 0$ is simply $\omega_p(-\gamma_p, \beta_p)$.  At these frequencies Eq. (\ref{conserv}) collapses to the relation 
$R_L R_R=1$, i.e. the reflection from one side is enchanced while from the other side is suppressed.  

Depending on the reflectance properties 
at $\omega_p$, we now distinguish two different types of perfect isolation: (a) In the first case, the reflection of an incident wave 
entering the sample in the direction that we have non-zero transmitivity, is zero. For example, the transport characteristics of a right incident 
wave would be  $T_{R}\neq 0$ and $R_{R} =0$ while, the same wave entering the circuit from the opposite (left) direction results in $T_{L}=0$ 
and $R_{L}\neq 0$. We refer to this case as {\it impedance-matched isolator} (IMI) as the transmitted signal does not experience any reflection 
at the TL-circuit interface. (b) In the second case the signal is completely absorbed (i.e. neither transmitted nor reflected) from one direction, 
while it is transmitted (and partially reflected) from the other.  We refer to this case as {\it unidirectional absorber} (UA). An example of such modes 
associated with the former case is marked with a (red) circle in  Fig. \ref{fig2}a while a representative mode for the latter case is indicated with 
a (blue) triangle in Fig. \ref{fig2}a.  Based on the symmetry relations of Eq. (\ref{symmetries}) we deduce that a change from a UA to a IMI behavior
occurs by simply performing the transformation $\beta\rightarrow -\beta$. Instead a simultaneous exchange of gain/loss elements i.e. 
$\gamma\rightarrow -\gamma$, will result in the same behavior (UA or IMI) but for an incident wave entering the structure from the opposite
direction.

In Fig.~\ref{fig2}b we report the left/right reflectances and transmitances, for a 1D network consisting 
of ${\cal N}$ identical EC units. We observe that a perfect isolation is now extended over a broad-band frequency window (see highlighted area). A 
simple theoretical understanding for the isolation action can be achieved by realizing that in this case $t^{\cal N}_L/t^{\cal N}_R =\det M_{\cal{N}}
=(\det{M})^{\cal{N}}$ where $M$ is the transfer matrix of each EC unit and $M_{\cal {N}}$ is the transfer matrix associated with the total array. A 
Taylor expansion of the right hand side around a frequency $\omega_p$ of an individual unidirectional unit requires knowledge of the $n-$th 
order derivatives at $\omega_p$ i.e. $\frac{d^n \det{M_{\cal{N}}}}{d \omega^n}\Big|_{\omega=\omega_p} = 0$ for $n =1,\cdots,{\cal{N}}-1$. 
These vanishing derivatives appreciably flatten the transmitance at a neighborhood around $\omega_p$, effectively creating perfect isolation for 
a broad-band frequency domain, in agreement with the simulations shown in Fig.~\ref{fig2}b. Moreover an IMI behavior is evident from our data.
We have also checked that the previous expectations associated with the $\beta, \gamma$-symmetries are confirmed in the case of 1D chains
as well. The same behavior is also present for the equivalent periodic photonic crystal consisting of ${\cal N}$ basic unidirectional units as can be
seen from Figs. \ref{fig2}c-e. 

Finally, we investigate the effect of imperfections along 1D chains with local ${\cal P{\tilde T}}$-symmetry. We assume that in the case 
of ECs the disorder is introduced on the capacitances $C$ which are independently distributed around a typical capacitance $C_0$ 
with a uniform probability distribution in the interval $[C_{1}; C_{2}]$. In the absence of any gain/loss elements, such 1D chains exhibit the phenomenon 
of Anderson localization. This phenomenon was 
predicted fifty years ago in the framework of quantum (electronic) waves by Anderson \cite{A58} and its existence has been confirmed in recent years 
by experiments with classical \cite{CSG00,HSPST08,LAPSMCS08,SBFS07} and matter waves \cite{A08,Ignuscio}. 

\begin{figure}
   \includegraphics[width=0.80\linewidth, angle=0]{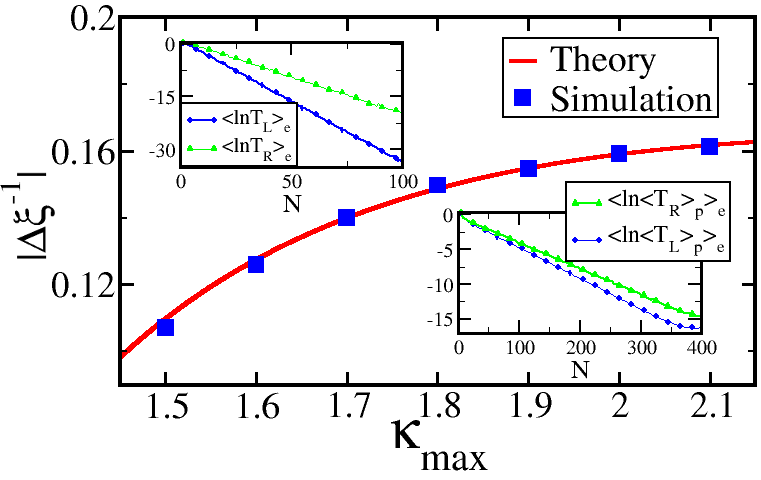}
    \caption{(Color online) Upper inset: The scaling of $\langle T_{L(R)}\rangle$ versus the system size ${\cal N}$ for a disordered 1D chain 
with random ${\cal P{\tilde T}}$-symmetric EC units. Main panel: The difference $\Delta \xi^{-1}$ between left and right localization lengths
for various $\kappa_2$ ($\kappa_1$ is kept fixed). The numerical results using transfer matrices (blue squares) are compared with the 
theoretical expression (solid line) of Eq. (\ref{diff}). Lower inset: The scaling of $\langle T_{L(R)}\rangle$ versus the system size ${\cal N}$ 
for the equivalent disordered photonic hetrostructure consisting of ${\cal P{\tilde T}}$-symmetric units.
\label{fig3}}
\end{figure}

Formally, the Anderson localization can be quantified via the so-called localization length $\xi$ which is defined as
\begin{equation}
\label{loclength}
\xi^{-1} \equiv -\lim_{{\cal N} \rightarrow \infty} {\langle \ln T \rangle \over {\cal N}}
\end{equation}
where $\langle \cdots \rangle$ denotes an ensemble average over disorder realizations. In reciprocal disordered systems, the localization length 
satisfies the relations $\xi_L = \xi_R$. We find instead that our set-up allows for {\it asymmetric localization} with $\xi_L\neq \xi_R$.  This is 
demonstrated in the upper inset of Fig. \ref{fig3} where we report the scaling behavior of $\langle \ln T_{L,R} \rangle$ for an 1D chain consisting of 
${\cal N}$ disordered EC units, each respecting locally a ${\cal P {\tilde T}}$-symmetry. Straightforward algebra leads to the following expression for 
the difference between the two localization lengths
\begin{eqnarray}
\label{diff}
\Delta\xi^{-1}=
\frac{2 
\biggl(a_{+}\ln |a_{+}|-a_{-}\ln|a_{-}| \biggr)\biggr|_{\kappa_{1}}^{\kappa_{2}}
}{(\kappa_{2} - \kappa_{1})\beta  (1-\mu^{2})\omega^{2}}
\end{eqnarray} 
where $\Delta\xi^{-1}\equiv\xi_L^{-1} - \xi_R^{-1}$, $a_{\pm}(\kappa) \equiv \beta(1 - \kappa(1-\mu^{2})\omega^{2}) \pm \gamma \mu$ and
$\kappa_{1,2}=C_{1,2}/C_0$.
In the main panel of Fig. \ref{fig3} we compare the theoretical predictions of Eq.~(\ref{diff}) together with the numerical results from the 
transfer matrix simulations of the disordered chain. 

The asymmetric localization $\xi_L \neq \xi_R$ can be also observed in the equivalent photonic crystal consisting of ${\cal N}$ random 
${\cal P{\tilde T}}$-symmetric units similar to the one shown in Fig. \ref{fig1}a. The disorder was introduced in the real part of the permitivity
of the active elements in such a way that in all cases the individual units were preserving a local ${\cal P{\tilde T}}$-symmetry. The numerical 
evaluation of $\langle \ln \langle T_{L,R}\rangle_p \rangle_e$ is performed using transfer matrices and the results are shown in the lower inset 
of Fig.~\ref{fig3}.  In this case, an additional average over different polarizations was performed in order to guarantee that the phenomenon is
polarization independent.

In conclusion, we have demonstrated both theoretically and numerically the dual behavior of  ${\cal P{\tilde T}}$-symmetric (photonic and electronic) 
circuits as perfect-impedance matched isolators or unidirectional absorbers. One-dimensional arrays consisting of such ${\cal P{\tilde T}}$-symmetric 
units can demonstrate these dual properties for a broad-band frequency window. Finally, we have shown that disordered 1D arrays with local 
${\cal P{\tilde T}}$-symmetry, show asymmetric (left/right) interference effects that can lead to asymmetric Anderson localization. It will be interesting 
to exploit the simplicity of ECs in order to investigate experimentally such phenomena and extend the analysis towards directional chaos and 
localization where a localization-delocalization phase transition was predicted even for 1D arrays \cite{HN96,E97,S98}. We expect  that fabricating 
such multi-functional components in the frame of electronic complementary metal oxide semiconductor (CMOS)  circuits \cite{volakis} and the 
frame of Indium Phosphide Photonic (InP)  monolithic Integrated Circuits \cite{bowers} will occur the years to come and the findings of the optics 
community will pollinate the silicon photonics and InP communities.

{\it Acknowledgements} The work identified in this paper was sponsored by the Air Force
Research Laboratory (AFRL/RY), through the Advanced Materials, Manufacturing and Testing Information Analysis Center (AMMTIAC) contract 
with Alion Science and Technology. Partial support from AFOSR grants No. FA 9550-10-1-0433 and LRIR 09RY04COR are also acknowledged. 


\vspace*{5cm}

\newpage

\textsc{Supplementary Material for Transmissions/Reflections amplitudes [Eq. (4)]}\\\\
%
%
%
The transmissions and reflections amplitudes can be expressed using the following three functions $f(\gamma,\beta)$, $g(\gamma,\beta)$, and $h(\gamma,\beta)$ which are defined for any frequency $\omega$ as
\begin{align*}
t_{R} = \frac{f(\gamma,\beta)}{h} ;\quad
t_{L} = \frac{f(-\gamma,\beta)}{h}; \quad
r_{L} = \frac{g(\gamma,\beta)}{ih} ;\quad
r_{R} = \frac{g(-\gamma,\beta)}{ih} \quad
\end{align*}
where $f$ and $g$ are real functions such that
\begin{align*}
f(\gamma,\beta) = f(-\gamma,-\beta) \qquad
g(\gamma,\beta) = g(\gamma,-\beta)
\end{align*}
%
%
%
To express these three functions the following shorthand notations are used:
\begin{align*}
M &\equiv (1-\mu^{2}) \\
K &\equiv M \left(\gamma ^2+\omega^2\right)-2\\\\
A &\equiv 2 M \left(M \left(\gamma ^2-2 \beta ^2\right)-2\right) \\
B &\equiv M \left(8 \beta ^2-4 \gamma ^2+\gamma ^4 M+2\right)+4 \\
C &\equiv 2 \left(\gamma ^2 \left(\mu ^2+1\right)-2 \beta ^2 M-2\right) \\
P &\equiv M^{2}\omega^{8} + A\omega^{6} + B\omega^{4} + C\omega^{2} + 1 \\\\
D &\equiv \eta ^2 \omega ^2 \left(M \omega ^2 K +1\right) \\
E &\equiv 2 \gamma  \eta  \omega ^2 \left(\mu ^2+M \omega ^2 K +1\right) \\
R &\equiv -2 \eta  \omega  \left(M \omega ^2-1\right) (\omega ^2 \left(
-2M \beta^2+K     )   +1 \right)
\end{align*}
Using the above expressions we get after straightforward algebra,
\begin{align*}
f(\gamma,\beta) &= 4 \beta  \eta  M \omega ^3 (\beta  M \omega^2 -\beta  +\gamma  \mu) \\
g(\gamma,\beta) &= E + P + D  \\
h &= R + i(P - D)
\end{align*}

\end{document}